# A New Method to Calculate Electromagnetic Impedance Matching Degree in One-Layer Microwave Absorbers*

MA Zhi(马治), CAO Chen-Tao(曹晨涛), LIU Qing-Fang(刘青芳),

WANG Jian-Bo(王建波)**

*Key Laboratory for Magnetism and Magnetic Materials of Ministry of Education,*

*Lanzhou University, Lanzhou 730000*

*Supported by the Fundamental Research Funds for the Central Universities under Grant No. lzujbky-2011-54, National Natural Science Foundation of China under Grant No. 11074101 and National Science Fund for Distinguished Young Scholars under Grant No. 50925103.

**Corresponding author. Email : wangjb@lzu.edu.cn; maz09@lzu.edu.cn.

Tel: 0931-8914171    Fax: 0931-8914160

A delta-function method was proposed to quantitatively evaluate the electromagnetic impedance matching degree. Measured electromagnetic parameters of α-Fe/Fe$_3$B/Y$_2$O$_3$ nanocomposites are applied to calculate the matching degree by the method. Compared with reflection loss and quarter-wave principle theory, the method accurately reveals the intrinsic mechanism of microwave transmission and reflection properties. A possible honeycomb structure with promising high-performance microwave absorption according to the method is also proposed.

*PACS*: 84.37.+q, 41.20.Jb, 77.22.Ch

In recent years, due to the increasingly serious electromagnetic (EM) interference caused by the rapid development of wireless communications and high frequency circuit devices in the gigahertz range, significant research efforts have been invested in the preparation of various structured magnetic (or dielectric) materials for the microwave absorption applications.[1-5] In particular, the magnetic or dielectric composites with high-performance microwave absorption properties have been intensively investigated.[6-13] It is found that strong microwave absorption of nanocomposites mainly results from proper EM impedance matching.[2, 14-17] The characteristic impedance of the absorbing material should be made equal to that of the free space to achieve zero-reflection. However, until now, there was no good method to accurately calculate the EM impedance matching degree due to too many parameters should be considered in designing the broad-band microwave absorbers.[18] To solve the above problem, we developed a delta-function method as an effective tool for evaluating the EM impedance matching degree. The new approach not only can provide a quantitative analysis of EM impedance matching degree but also can give guide to design microwave absorbing materials with better performance in a broad frequency range.

Figure 1(a) shows the toroidal samples (7.0 mm outside diameter and 3.04 mm inner diameter) which are usually used in measuring the scattering parameters. The real and imaginary part of the relative complex permittivity/permeability are usually determined from the complex scattering parameters using the Nicolson–Ross model by the vector network analyzer [19]

$$\varepsilon_r = (1/\varepsilon_0) \bullet (\varepsilon' - j\varepsilon'') = (\varepsilon'/\varepsilon_0) \bullet (1 - j\tan\delta_e) \tag{1}$$

$$\mu_r = (1/\mu_0) \bullet (\mu' - j\mu'') = (\mu'/\mu_0) \bullet (1 - j\tan\delta_m) \tag{2}$$

Where $\varepsilon_r$ is the relative complex permittivity, $\mu_r$ is the relative complex permeability. Tan$\delta_e$ is dielectric dissipation factor, tan$\delta_m$ is magnetic dissipation factor. The dependences of microwave absorption on the frequency, thickness and the relative EM parameters are obtained based on one-layer microwave absorbers model [20-21] (see Fig. 1(b)). The input impedance $Z_{in}$ at the air–material interface was calculated from measured complex permittivity ($\varepsilon_r$) and permeability ($\mu_r$) at the given thickness based on transmission line theory [22]

$$Z_{in} = Z_0\sqrt{\mu_r/\varepsilon_r}\tanh(j2\pi fd\sqrt{\mu_r \bullet \varepsilon_r}/c) \tag{3}$$

Where $Z_0$=377 Ω is the intrinsic impedance of free space, $f$ is the microwave frequency, $c$ is the speed of light and $d$ is the thickness of the absorbers. The reflection loss (*RL*) of absorbers backed with a perfect conductor in the one-layer model can be written as [20, 22]

$$RL(dB) = 20\log_{10}|(Z_{in} - Z_0)/(Z_{in} + Z_0)| \tag{4}$$

According to the above equations, the *RL* value was implicitly determined by $d$, $f$, $\mu_r$ and $\varepsilon_r$. The dip in *RL* indicates the microwave absorbing power of the materials. When $Z_{in}$ approaches $Z_0$, RL gradually approaches -∞. The condition that an EM wave is completely absorbed by the absorbing medium on the ideal conductor could be written as:

$$1/\tanh(j2\pi fd\sqrt{\mu_r \bullet \varepsilon_r}/c) = \sqrt{\mu_r/\varepsilon_r} \tag{5}$$

Equation (5) is a complex function (both the left and the right side of the equation should have a real part and an imaginary part). To obtain real function, real and imaginary parts of the relative permittivity and permeability are introduced to equation (5).[18] According to the trigonometric function and complex equation theory, finally, the following equations could be derived from the equation (5):

$$\begin{cases} \cos\theta = \dfrac{\mu'\cos\delta_e - \varepsilon'\cos\delta_m}{\mu'\cos\delta_e + \varepsilon'\cos\delta_m} \times \cosh(\dfrac{4\pi fd\sqrt{\mu'\varepsilon'}\sin[(\delta_e+\delta_m)/2]}{c\sqrt{\cos\delta_e \cos\delta_m}}) \\ \sin\theta = \tan\dfrac{\delta_e+\delta_m}{2}\sinh(\dfrac{4\pi fd\sqrt{\mu'\varepsilon'}\sin[(\delta_e+\delta_m)/2]}{c\sqrt{\cos\delta_e \cos\delta_m}}) \end{cases} \tag{6}$$

Where $\theta$ could be presented by

$$\theta = \dfrac{4\pi fd}{c} \bullet \sqrt{\mu'\varepsilon'} \bullet \dfrac{\cos(\delta_e/2+\delta_m/2)}{\sqrt{\cos\delta_e \cos\delta_m}} \tag{7}$$

The equation (6) is equivalent to the complex function (5). In order to evaluate the EM impedance matching degree, the above two equations should be revised into only one equation. According to $\sin^2\theta + \cos^2\theta = 1$, equation (6) could be rewritten and simplified as the following simple equation: $\sinh^2(Kfd) = M$, where $K$ and $M$ could be determined by the relative complex permittivity and permeability:

$$K = 4\pi\sqrt{\mu'\varepsilon'}/c \times \dfrac{\sin[(\delta_e+\delta_m)/2]}{\cos\delta_e \bullet \cos\delta_m} \tag{8}$$

$$M = \frac{4\mu'\cos\delta_e \varepsilon'\cos\delta_m}{(\mu'\cos\delta_e - \varepsilon'\cos\delta_m)^2 + (\tan\frac{\delta_m - \delta_e}{2})^2(\mu'\cos\delta_e + \varepsilon'\cos\delta_m)^2} \quad (9)$$

Now, the EM impedance matching degree could be described as a delta-function $\Delta = |\sinh^2(Kfd) - M|$. Here, the delta-function represents the impedance matching degree. It means that the higher absorption (where *RL* approaches -∞) at a certain thickness is the result of better EM impedance matching, where delta value should gradually approaches zero. The smaller delta value implies better EM impedance matching. By the method, once the permittivity and permeability were obtained from the experiment, like measured by the vector network analyzer, the EM impedance matching degree could be accurately evaluated and quantificationally analyzed by the delta values.

In order to judge the effectiveness of the method and draw a comparison, real EM parameters measured in real materials are applied to calculate the matching degree. Since permittivity/permeability of α-Fe/Fe$_3$B/Y$_2$O$_3$ nanocomposites showed almost constant over the plotted frequency range (1-10 GHz, $\varepsilon_r$=15-0.6j, $\mu_r$=1.25-0.5j),[23] the data are used in latter calculations. We hypothesized that the permittivity and permeability were maintain constant over whole 0.1-18 GHz range. The calculated reflection loss and delta values are shown in Fig. 2(a) and (b). It can be seen that reflection loss and delta values show same tendency. The comparison of these two maps concludes that where the reflection loss below -8 dB, the delta values should be smaller than 0.2. It is real that the delta values well present the EM impedance matching degree and the method is effective.

In Fig. 2(a) and (b), it can further be noticed that the minimum *RL* and delta values shift towards a higher frequency side if the sample thickness is reduced. This can be understood based on quarter-wave principle[1], which is written as: $d = (n/4)c/[f(|\mu_r||\varepsilon_r|)^{1/2}]$. Where *n* is odd number, $|\mu_r|$ and $|\varepsilon_r|$ are the module of $\mu_r$ and $\varepsilon_r$, respectively. Since the thickness *d* is inversely proportional to frequency *f*, the above criterion is satisfied at reduced thickness for higher frequencies. The higher absorption at reduced thickness is the result of total cancellation, where *RL* and delta values should also be reduced, satisfying the above criteria perfectly. Fig. 2(c) shows the matching thickness and matching frequencies based on the above quarter-wave principle. The filled regions implied the ones where the reflection loss or the delta-function values reaches their minimum values. From the comparison in Fig. 2(a) to (c), it is found that the reflection loss was crucially determined by the quarter-wave principle at the given dielectric and magnetic parameters. In those corresponding regions, the EM impedance matching was well achieved, when the reflection loss below -8 dB, the delta values should be smaller than 0.2 and *n* should be equal to 9.

The above analysis also implies that the good microwave absorption property of the magnetic materials was mainly a result of structure effect when the materials are certain. In other words, when EM impedance of absorbers was much better matched over a broad-band frequency range, the thickness of absorbing materials should smoothly decrease as frequency continuous increases. So we think that the structure of the one-layer microwave absorber should be not uniform but have a proper gradient. The bottom graph in Fig. 3 shows one possible honeycomb structure of one-layer microwave absorbers. Based on the former analysis, the curvature of inner surface of the honeycomb structure should satisfy some requirements. The top right graph in Fig. 3 illustrates the section of the honeycomb structures and the red box shows an element of the honeycomb structures. The top left graph implies that element of the honeycomb structure should have an optimum gradient, which should be designed to fully satisfy the delta function or approximately satisfy the quarter-wave principle. It is promising that this optimum honeycomb structure will have a better microwave absorption performance. Furthermore, other structures also could be designed according to this view for anti-EM interference technology induced by the wireless communication equipment, since most of the wireless communication equipments are working in 2–4 GHz. The structures also could be adapted to the radar wave absorption region, since the working frequency of most radar is in 8–18 GHz.

In conclusion, a delta-function method was proposed to calculate the EM impedance matching degree. It is demonstrated that this method is useful to analyze the EM impedance matching degree and can be used to design lightweight high-performance microwave absorbers. Based on this function method, it is easier to obtain the optimum gradient curves for designing the honeycomb-shaped or other shaped microwave absorbing materials. The results in this work are helpful for investigating the optimum structures of microwave absorber in practice and would benefit the design of lightweight high-performance microwave absorbers in the future.

**References**

[1] Yusoff A N, Abdullah M H, Ahmad S H, Jusoh S F, Mansor A A and Hamid S A A 2002 *J. Appl. Phys.* **92** 876

[2] Li Z W, Lin G Q and Kong L B 2008 *IEEE Trans. Magn.* **44** 2255

[3] Sakai K, Asano N, Wada Y and Yoshikado S 2010 *J. Eur. Ceram. Soc.* **30** 347

[4] Nie Y, He H H, Gong R Z and Zhang X C 2007 *J. Magn. Magn. Mater.* **310** 13

[5] Yi H B, Wen F S, Qiao L and Li F S 2009 *J. Appl. Phys.* **106** 103922

[6] Liu X G, Li B, Geng D Y, Cui W B, Yang F, Xie Z G, et al. 2009 *Carbon* **47** 470

[7] Cao M S, Song W L, Hou Z L, Wen B and Yuan J 2010 *Carbon* **48** 788

[8] Dong X L, Zhang X F, Huang H and Zuo F 2008 *Appl. Phys. Lett.* **92** 013127

[9] Li B W, Shen Y, Yue Z X and Nan C W 2007 *J. Magn. Magn. Mater.* **313** 322

[10] Wu M Z, He H H, Zhao Z S and Yao X 2000 *J. Phys. D: Appl. Phys.* **33** 2398

[11] Yan S J, Zhen L, Xu C Y, Jiang J T and Shao W Z 2010 *J. Phys. D: Appl. Phys.* **43** 245003

[12] Yan L G, Wang J B, Han X H, Ren Y, Liu Q F and Li F S 2010 *Nanotechnol.* **21** 095708

[13] Xi L, Wang Z, Zuo Y L and Shi X N 2011 *Nanotechnol.* **22** 045707

[14] Liu X G, Geng D Y, Meng H, Shang P J and Zhang Z D 2008 *Appl. Phys. Lett.* **92** 173117

[15] Kang Y Q, Cao M S, Yuan J and Shi X L 2009 *Mater. Lett.* **63** 1344

[16] Yan S J, Zhen L, Xu C Y, Jiang J T, Shao W Z and Tang J K 2011 *J. Magn. Magn. Mater.* **323** 515

[17] Han R, Han X H, Qiao L, Wang T and Li F S 2011 *Physica B* **406** 1932

[18] Huang Y Q, Yuan J, Song W L, Wen B, Fang X Y and Cao M S 2010 *Chin. Phys. Lett.* **27** 027702

[19] Nicolson A M and Ross G F 1970 *IEEE Trans. Instrum. Meas.* **19** 377

[20] Kim S S, Jo S B, Choi K K, Kim J M and Churn K S 1991 *IEEE Trans. Magn.* **MAG-27** 5462

[21] Shin J Y and Oh J H 1993 *IEEE Trans. Magn.* **29** 3437

[22] Kwon H J, Shin J Y and Oh J H 1994 *J. Appl. Phys.* **75** 6109

[23] Liu J R, Itoh M and Machida K 2003 *Appl. Phys. Lett.* **83** 4017

**Figure captions:**

**Fig. 1** (a) Coaxial sample used in measuring the complex scattering parameters, and (b) one layer microwave absorbers model was used to calculate the electromagnetic impedance matches degree.

**Fig. 2** (a) Calculated microwave reflection loss map by equation (4), (b) calculated delta values map and (c) calculated quarter-wave principle map. The relative permittivity and permeability were hypothesized at constant over 0.1-18 GHz ($\varepsilon_r$=15-0.6j, $\mu_r$=1.25-0.5j).

**Fig. 3** Optimum honeycomb structures were designed for high-performance microwave absorbers. The bottom graph gives the 3D honeycomb structures. The top right graph shows the section of the perfect honeycomb structures and the top left graph shows the gradient of the honeycomb element should satisfy the delta function.

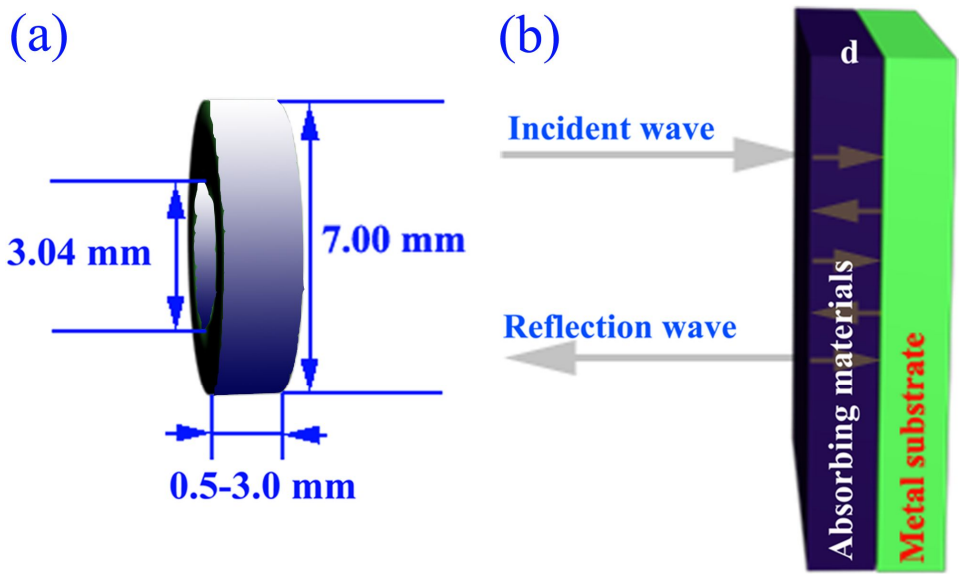

Fig. 1.

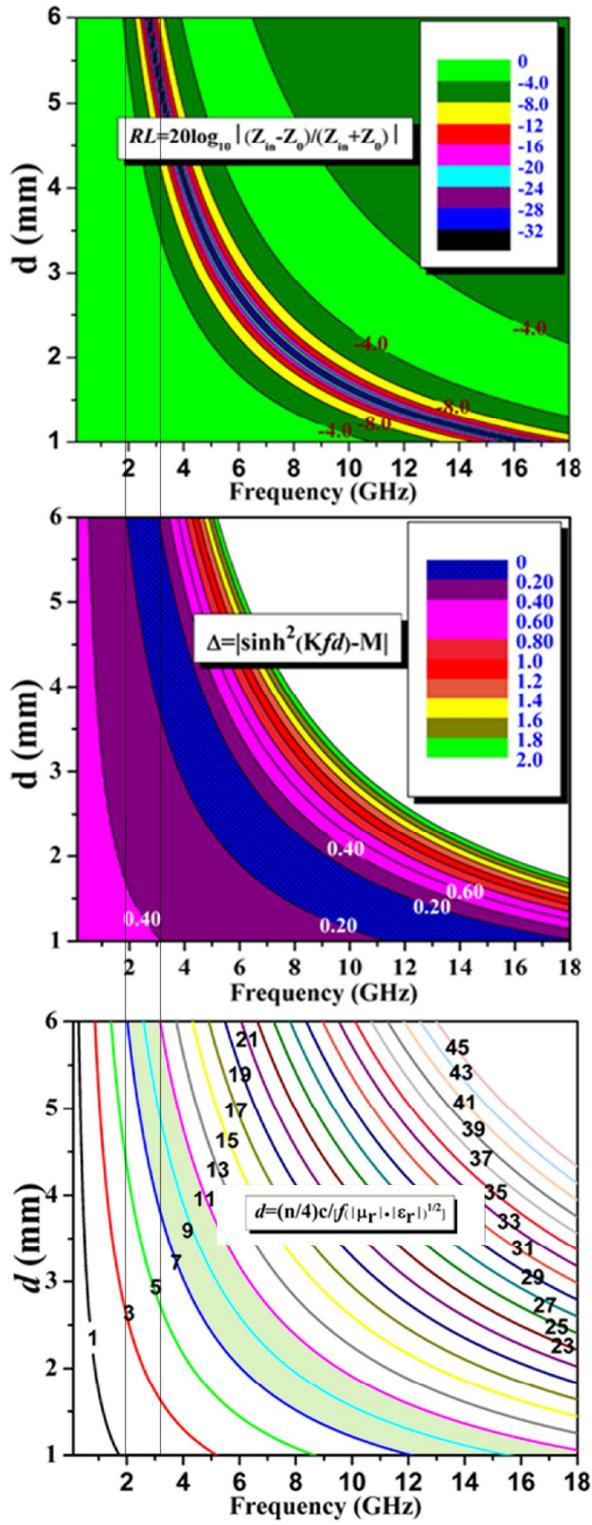

Fig. 2.

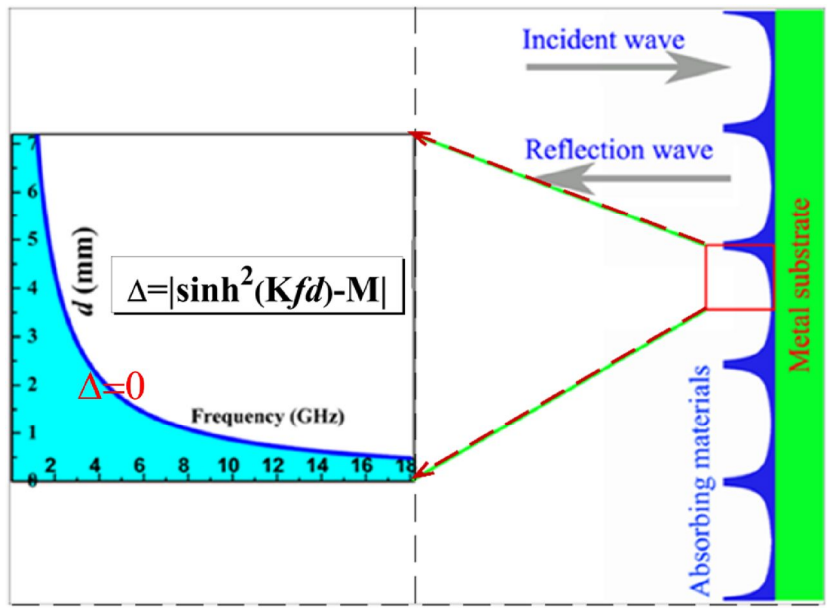

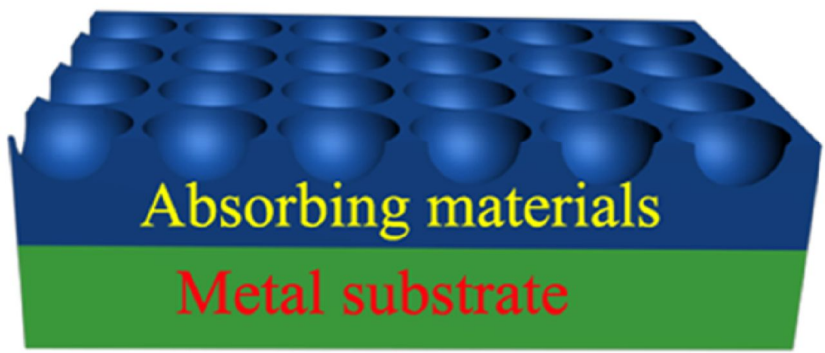

Fig. 3.